\documentclass[apj]{emulateapj}
\usepackage{gensymb}
\usepackage{amsmath}
\usepackage{graphicx}
\newcommand{\snr}[0]{G16.05-0.57}

\begin{document}
\title{Classifying structures in the ISM with Support Vector Machines: \\ the G16.05-0.57 supernova remnant }
\shortauthors{Beaumont, Williams \& Goodman}
\author{Christopher N. Beaumont $^{1,2}$,
Jonathan P. Williams$^1$, Alyssa A. Goodman$^2$}
\affil{$^1$Institute for Astronomy, University of Hawai'i, 2680 Woodlawn Drive, Honolulu HI 96822; beaumont@ifa.hawaii.edu}
\affil{$^2$Harvard-Smithsonian Center for Astrophysics, 	60 Garden St., Cambridge MA 02138}
\shorttitle{Automated Classification of the ISM}
\slugcomment{ApJ in Press (September 2011, v738-2)}

\begin{abstract}
We apply Support Vector Machines -- a machine learning algorithm --  to the task of classifying structures in the Interstellar Medium. As a case study, we present a position-position velocity data cube of $^{12}$CO J=3--2 emission towards \snr{}, a supernova remnant that lies behind the M17 molecular cloud. Despite the fact that these two objects partially overlap in position-position-velocity space, the two structures can easily be distinguished by eye based on their distinct morphologies. The Support Vector Machine algorithm is able to infer these morphological distinctions, and associate individual pixels with each object at $>$90\% accuracy. This case study suggests that similar techniques may be applicable to classifying other structures in the ISM -- a task that has thus far proven difficult to automate. 
\end{abstract}

\maketitle

\section{Introduction}
\label{sec:intro}

Classifying interesting objects in a dataset is an essential early step in most analysis tasks. For many objects in astronomy (stars, galaxies, solar system objects, extragalactic supernovae), algorithms can identify and characterize these objects automatically \citep{Irwin85, Bertin96, Naylor98}. Structures in the Interstellar Medium (ISM), however, have proven difficult to classify in this way. These objects -- which include, e.g., molecular clouds, infrared dark clouds, bubbles, jets, radiation-shaped pillars, and filaments -- are morphologically complex and heterogeneous. The essential properties of these structures are hard to encode.  

To take advantage of increasingly wide-area surveys, previous researchers have mainly relied on manual identification of features in the ISM \citep{Churchwell06, Helfand06, Curtis10, Arce10}. There are several drawbacks to this approach: it is time consuming, non-repeatable and, when identifying complex structures, affected by difficult-to-quantify selection effects related to how specific people perceive an image. Despite recent advances in the study of specific objects (e.g. Infrared Dark Clouds, \citealt{Peretto09};  filaments, \citealt{Menshchikov10}), automated feature identification in the ISM is an still an open problem.

Machine learning algorithms are designed to infer patterns in data which are otherwise difficult to define explicitly. They are grouped into two classes: supervised (in which the algorithm is ``trained'' to recognize a pattern via a set of training examples) and unsupervised (in which the algorithm identifies groupings within a dataset \textit{a priori}). 

These algorithms can mechanize the process of object identification, and hence address many of the shortcomings of manual  classification -- they scale easily to other similar data, and their results are repeatable. A potential drawback of these methods is that, because the classification is not guided by a physical model, they are susceptible to over-fitting and to inheriting selection biases within the training data. However, because the machine approach extends easily to other similar data, these biases are more readily characterized via classification of test data sets.

In this paper, we explore whether machine learning algorithms can be used to catalog structures in the ISM. We present as a case study a spectral line data cube of $^{12}$CO emission towards the M17 star forming region. Emission from this cloud overlaps with \snr{}, a supernova remnant situated behind the cloud. We use the Support Vector Machine (SVM, \citealt{Vapnik95}) algorithm to identify the supernova remnant and, on a pixel-by-pixel basis, classify the original data cube. We are able to use this classification to derive the mass and momentum of the supernova remnant, which would not otherwise be possible with these data.

\section{The data}

\begin{figure*}
\begin{center}
\includegraphics[width=3in]{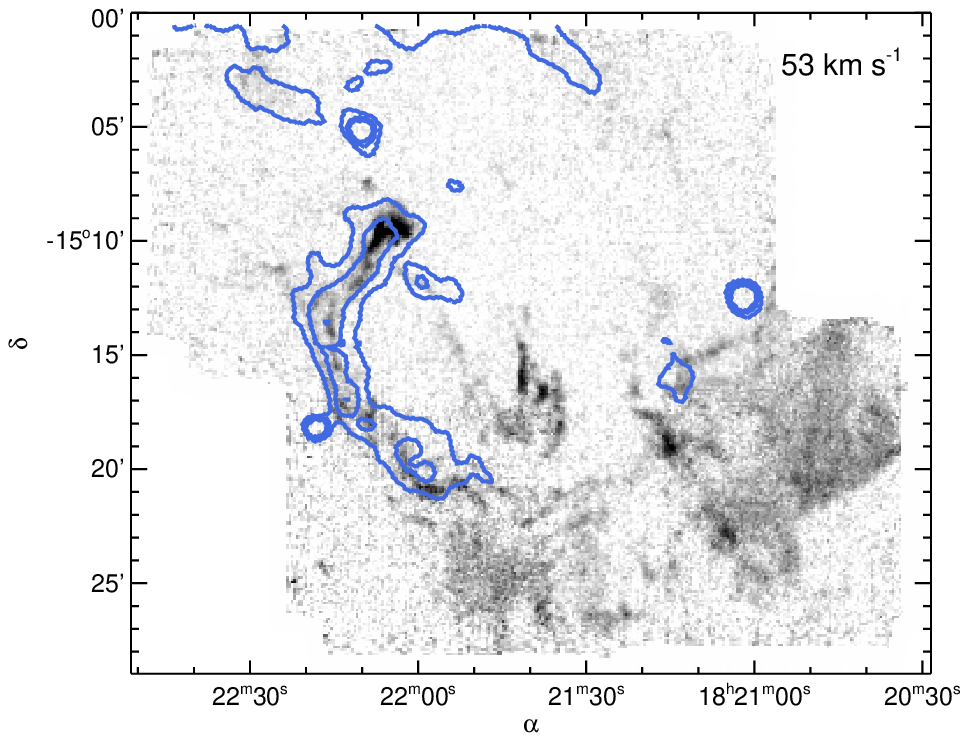}
\includegraphics[width=3in]{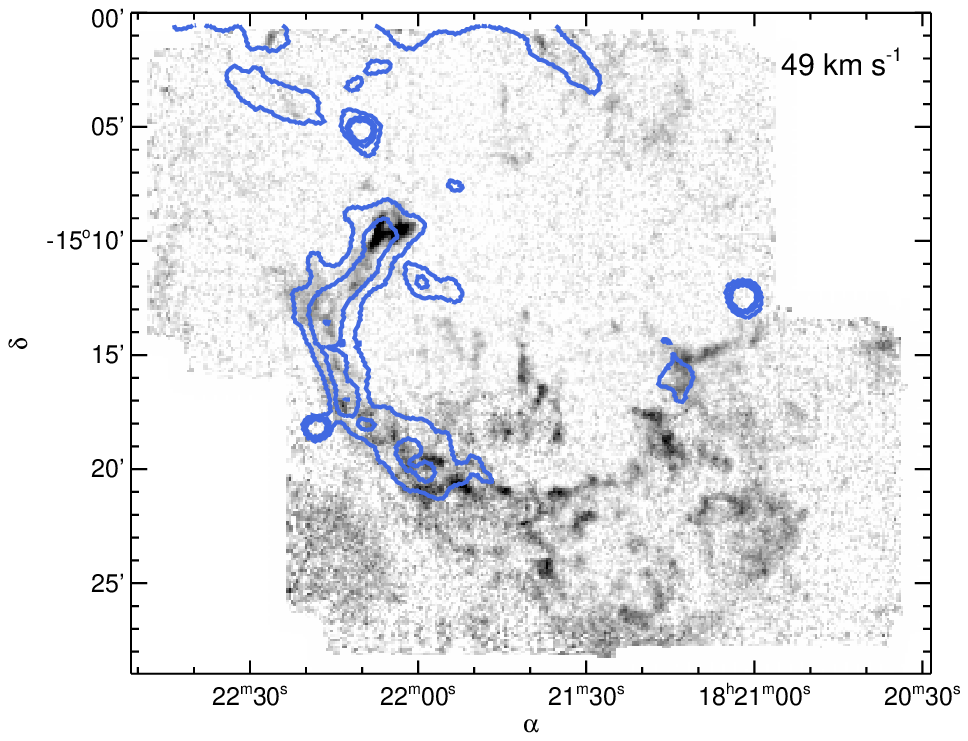}
\includegraphics[width=3in]{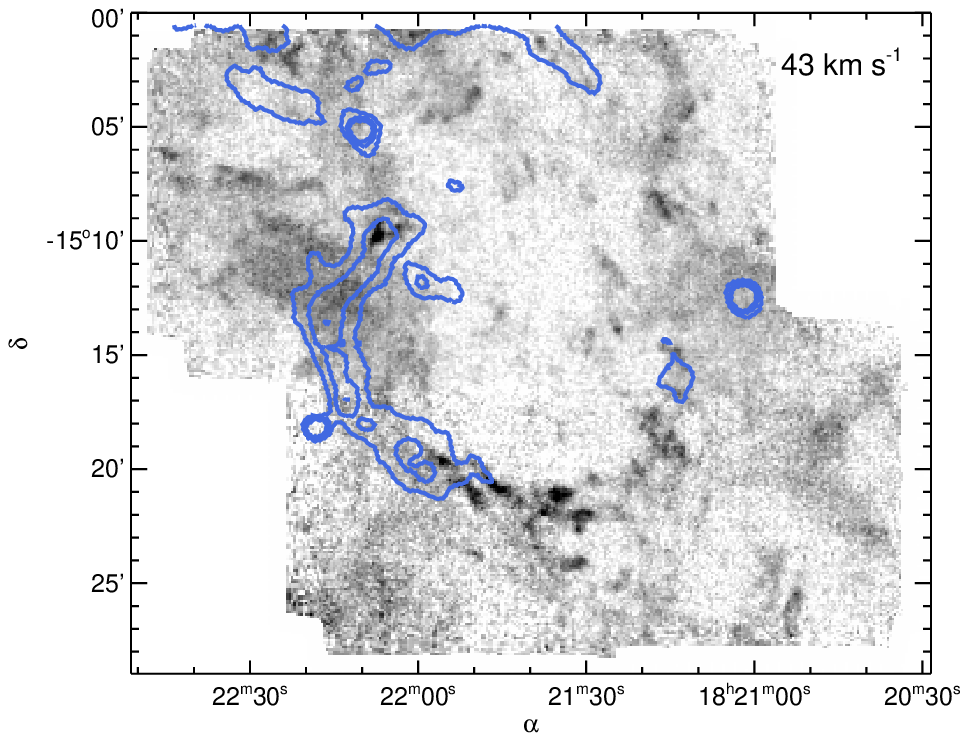}
\includegraphics[width=3in]{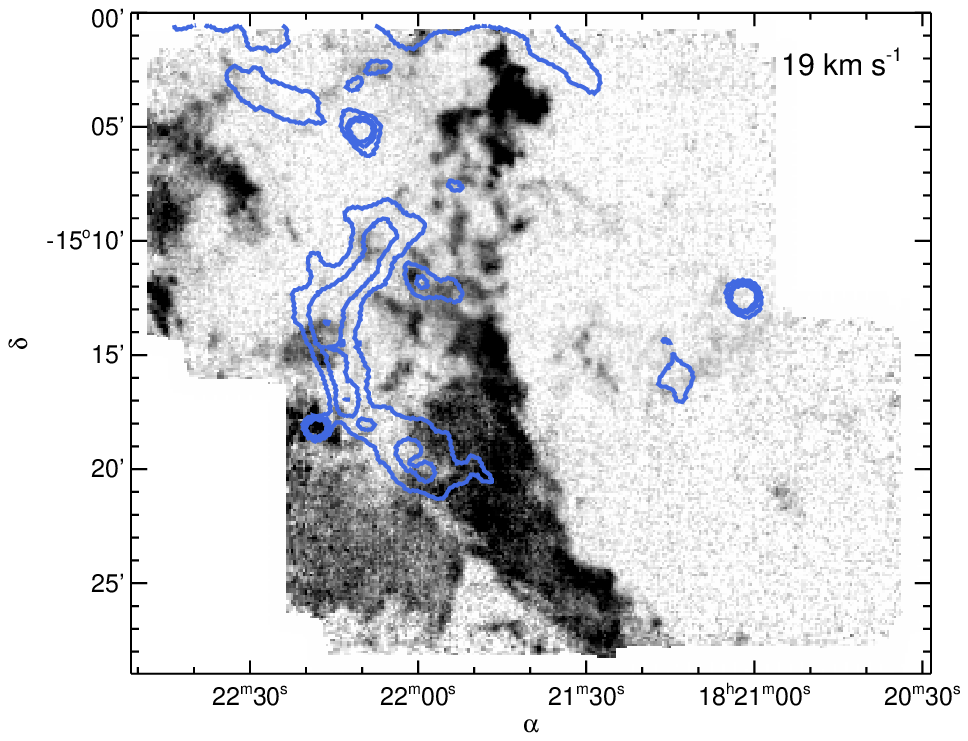}
\end{center}
\caption{The \snr{} supernova remnant. The greyscale shows the intensity of our CO J=3--2 map at several velocities, while the blue contours trace the 20\,cm continuum emission presented in Brogan et al (2006). The contour lines are at 1, 1.5, and 2$\sigma$ above the background. The compact, filamentary emission especially noticeable at 49\,km\,s$^{-1}$ is from the supernova remnant. The diffuse emission in the first three panels is from clouds in the Scutum arm, while the bright emission at 19\,km \,s$^{-1}$ is part of the M17 molecular cloud.}
\label{fig:brogan}
\end{figure*}

\snr{} is a supernova remnant in the inner galaxy first discovered via its cm synchrotron emission by Brogan et al. (\citeyear{brogan06}). Because it is located behind M17, we serendipitously observed the remnant with the James Clerk Maxwell Telescope (JCMT) during an imaging campaign of the latter region. Our observations of \snr{} were taken on the nights of 2009 June 23-25, using the HARP heterodyne receiver array \citep{Smith08}. The observations target the $^{12}$CO J=3--2 line, which traces moderately excited ($h\nu / k=16$K) and dense ($n_{\rm crit} \sim 10^3 ~{\rm cm}^{-3}$) gas at a resolution of $15''$. The data were acquired via position-switched raster scans, using a reference position of ($\ell, b$)=($15.5\degree, -2.4\degree$). Basket-weaving was used to reduce striping artifacts in the final map (see, e.g., Section 2.1 of \citealt{Davis10}). Weather conditions were grade 2-3 (225 GHz opacity $\tau_{225}$ = 0.06--0.10). To convert from antenna temperature $T_A^*$ to main beam temperature ($T_{mb} = T_R^* = T_A^* / \eta$) we assumed an aperture efficiency of $\eta = 0.63$ \citep{Buckle09}

\snr{} is visible as a filamentary arc subtending 20 arcminutes on the sky (Figure \ref{fig:brogan}). To our knowledge, these are the first molecular line observations of this object. Features belonging to the supernova have typical linewidths of 15 km/s FWHM, and are centered on a wide range of velocities from -5 -- 90 km/s. It is impossible to derive a kinematic distance from such broad and scattered lines, but supernova features are absorbed by overlapping cloud emission at 40 km/s. This corresponds to a near kinematic distance of 3.4 kpc, coincident with the Scutum galactic arm. The fact that \snr{} emits CO line emission suggests that the remnant is in the process of colliding with another molecular cloud, and we are observing the shocked interface of this collision \citep{Scoville77, vanDishoeck93}.

\begin{figure}
\includegraphics[width=3.5in]{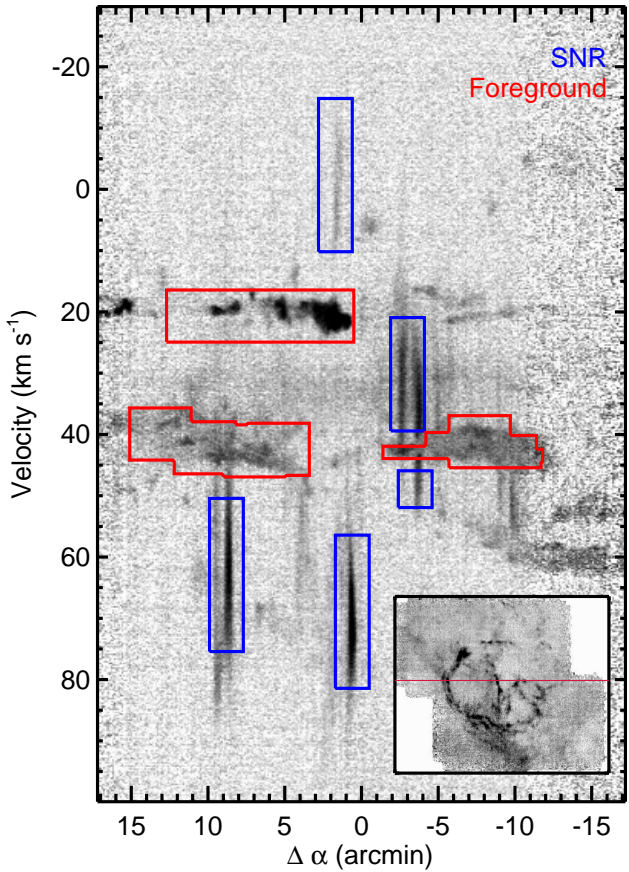}
\caption{A position-velocity slice through the data cube, manually labeling some of the emission from the supernova remnant (blue) and foreground clouds (red). The inset shows a position-position view of the cube, and the red line shows the location of the PV slice. The bright feature at 20 km/s is the M17 cloud, while the emission at 40 km/s is from the Scutum Arm. Individual supernova emission features are distinguished by their narrow spatial extent and broad velocity profiles.  Shocked emission from the expanding supernova is detected at all velocities throughout the cube. As a result, it overlaps other structures in several places.}
\label{fig:slice} 
\end{figure}

The morphology of structures in this datacube make for an appropriate case study of automated feature classification in molecular line datasets. A representative position-velocity slice through the data cube (Figure \ref{fig:slice}) shows that the remnant overlaps molecular cloud emission from M17 and the Scutum arm. However, the supernova has markedly different structure; features from the remnant are spatially compact and extend over tens of km/s in velocity space, while the other structures it overlaps with are more spatially diffuse and kinematically narrow. This suggests that a learning algorithm may be able to  distinguish emission from \snr{} based on its unique morphology. 

\section{Support Vector Machines}
The Support Vector Machine algorithm is a supervised learning algorithm which attempts to segregate data points into two categories, based on a representative sample of data belonging to each category. The method has recently been applied to many diverse problems in astronomy, including redshift estimation \citep{Wang08}, galaxy morphology identification \citep{Huertas08}, and time series analysis \citep{Kim11}. Here we provide a basic overview of the algorithm, but refer the reader to Press et al. (\citeyear{Press07}) for a deeper and more precise derivation, and to Vapnik (\citeyear{Vapnik99}) for a discussion of the algorithm's foundation in statistical learning theory. In what follows, we use the SVM$^{light}$ implementation by Joachims (\citeyear{Joachims99}). While this code is written in C, we have written a set of wrappers to use these tools within IDL.

During training, the SVM algorithm takes as input a feature vector for each training example -- a set of $N$ quantities that describe the discriminating properties of that object. These numbers can be thought of as coordinates for a vector in $N$-dimensional feature space. Using a training set of pre-classified feature vectors, the algorithm searches for a decision boundary in feature space that optimally separates examples from each category. New data points are then assigned a classification based on the side of the decision boundary on which they fall.

More specifically, SVM seeks a decision boundary $B$ that maximizes a fitness function $F$ given by
\begin{equation}
F =M - C ~\sum_i{\xi_i(B, M)} 
\label{eq:svm}
\end{equation}

Here $M$ is the margin of the boundary; SVM attempts to separate training data with a large gap between each data point and the boundary, and $M$ defines the size of this gap. $\xi_i(B, M)$ is the degree to which training example $i$ violates this criterion. If example $i$ is separated by more than $M$ from $B$ (and on the correct side), $\xi_i=0$. Otherwise, $\xi_i$ is the distance that $i$ would have to be moved to satisfy this condition. The adjustable cost parameter $C$ sets a tradeoff between large margins $M$ and poor classifications $\xi_i$. 

\begin{figure}
\includegraphics[height=2in]{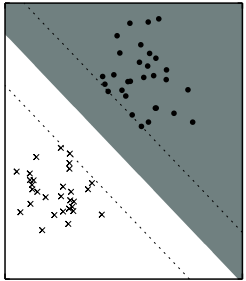}
\caption{A simple classification task in two dimensions. Dots and Xs represent training examples from two classes. Each training example consists of a 2-dimensional feature vector, represented in the figure by the position on the plane. The grey and white areas denote which regions of feature-space an SVM classifier assigns to each class. The SVM algorithm seeks a decision boundary which maximizes the distance between training data and the boundary (the margin, shown here as dotted lines).}
\label{fig:svm}
\end{figure}

For a better understanding of Equation \ref{eq:svm}, consider first the case where $C$ is very large, the feature space is 2-dimensional, and the decision boundary is  restricted to a line (Figure \ref{fig:svm}, where the different symbols denote training examples from two different classes). In this scenario, the algorithm first optimizes Equation \ref{eq:svm} over $M$ for a fixed boundary. Since $C$ is very large, even a single training example on the wrong side of the margin will heavily penalize Equation \ref{eq:svm}. Thus, the optimal $M$ will be the margin that just touches the training example closest to $B$ (i.e., the largest M that satisfies $\sum_i{\xi_i(B,M)} = 0$). The algorithm repeats this optimization over all boundaries, finding the plane that can accommodate the largest margin. The final classification is illustrated in Figure \ref{fig:svm}; the background shading shows how the algorithm classifies feature space. The dotted line traces the margin on either side of the boundary.

\begin{figure}
\includegraphics[height=1.4in]{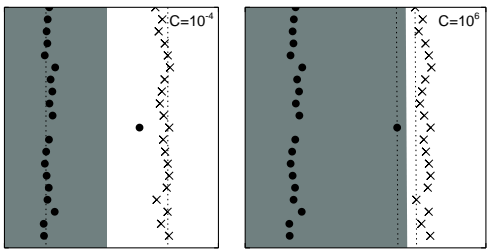}
\caption{An illustration of how the cost factor $C$ influences the SVM classification. The symbols and colors are the same as in Figure \ref{fig:svm}. The left and right figures show classifications with low and high values of $C$. Decreasing $C$ decreases the penalty for mis-classifications near the boundary, increasing robustness against outliers. }
\label{fig:svm_c}
\end{figure}

Next, consider the impact of reducing $C$. Individual $\xi_i$ terms now penalize Equation \ref{eq:svm} less heavily. The optimal boundary in this scenario may be one which misclassifies a small number of outliers, but can afford to partition the remaining data with a larger margin $M$. This is depicted in Figure \ref{fig:svm_c}.

\begin{figure}
\includegraphics[width=3.4in]{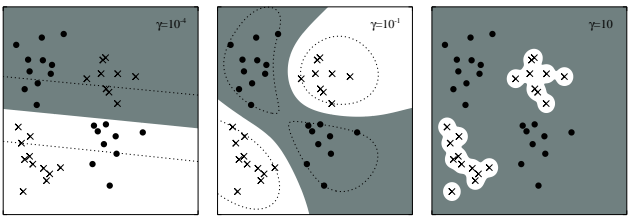}
\caption{An illustration of how the RBF kernel parameter $\gamma$ influences the SVM classification. As $\gamma$ increases, the decision boundary becomes more curved, and conforms more tightly to individual training examples. The margin in the rightmost plot wraps very tightly around the Xs, and is not drawn.} 
\label{fig:svm_gamma}
\end{figure}

In addition to $C$, a second set of adjustable parameters characterize a ``kernel function''. The kernel function determines the topology of the decision surface. In the simplest case, decision boundaries are hyper-planes in feature space. In this paper, we use the radial basis kernel function (RBF), a popular and effective kernel that allows for non-planar decision boundaries. The RBF kernel has one adjustable parameter, $\gamma$, that controls the curvature of the decision surface. For example, Figure \ref{fig:svm_gamma} shows an SVM classification of a data set with three different values of $\gamma$. Low values lead to stiff boundaries, while high values lead to curved boundaries that may over-fit to the training data. In this application, over-fitting refers to the case when the decision boundary conforms too tightly to the individual training examples, and the larger-scale organization of the data is ignored. 

A small complication arises when different elements of the feature vector have different scales. Components of the feature vector with the very large numerical values will dominate $\xi$, and the remaining components will have a negligible effect on the classification. To circumvent this, we normalize each element of the feature vectors, such that the dispersions of each element across the data are equal. 

From a practical standpoint, then, an SVM-based classification task involves four steps:
\begin{enumerate}
\item Manually classify a representative subset of the data (\S \ref{sec:train}).
\item For each classified example, create a feature vector. This vector should encode the properties of an object that make it identifiable (\S \ref{sec:feature}).
\item Choose a kernel function. In this study, we restrict our attention to the radial basis function.
\item Train the algorithm, and optimize the classification by adjusting free parameters ($C$ and $\gamma$ in this case; \S \ref{sec:optimize}). 
\end{enumerate}

In what follows we apply these steps to disentangle overlapping emission from the supernova remnant \snr{} and foreground molecular clouds.

\subsection{Manually classifying data}
\label{sec:train}

The success of any supervised learning algorithm is limited by how representative the training set is. Because our aim is a pixel-by-pixel classification of the data, our training set consists of a subset of these pixels, manually-classified as associated with the SNR (or not). We carried out this manual identification on subsets of four position-velocity slices through the cube. The classification of one of the planes is shown in Figure \ref{fig:slice}. In total, the training set explicitly labels $0.4\%$ of the pixels in the cube (5\% of the pixels above $3 \sigma$). However, as we show in Section \ref{sec:performance}, only a small fraction (~5\%) of these examples are ultimately necessary.

\subsection{Creating a feature vector}
\label{sec:feature}
Each pixel in our training set must be assigned a feature vector -- a list of numerical attributes that distinguish between the supernova and unassociated foreground objects. When classifying a pixel in the data by eye, it is sufficient to examine the pixels in the immediate vicinity. In particular, a $30\times30\times100$ pixel sub-cube  in PPV space is sufficient for a human to classify the pixel in the center of that cube. At the presumed 3.4 kpc distance to the supernova, this corresponds to a $3 {\rm pc} \times 3 {\rm pc} \times 25 {\rm ~km ~s}^{-1}$ region. In principle, one could use the intensities of these $9 \times 10^4$ pixels as a feature vector for the central point. In practice, such a large feature vector is prohibitively slow. We tested three strategies to compress this information. Each of these strategies defines a different feature vector:

\textbf{Moment}. For each pixel $p_i$, we extract the surrounding $30\times30\times100$ pixels. We calculate the mean intensity of this cube, and the first and second moments along each direction through the data. These seven numbers constitute the feature vector for $p_i$.  Relative to other cloud emission, supernova features have large velocity dispersions and small spatial dispersions -- this information is encoded in the moments of the data. 

\textbf{Derivative}. Spatial derivatives are sensitive to edges in images, and such information can be used to identify filamentary structures in astronomical data \citep{Molinari10}. We generate a feature vector that encodes this information. We approximate the gradient in each direction and pixel location using the Sobel edge detection operator. To generate the feature vector for pixel $p_i = (x_0, y_0, z_0)$, we sample profiles of each derivative along each direction through the pixel:
\begin{align}
\nonumber P_1 &= \left \{ \partial_x(x_0 + \delta, y_0, z_0) ~|  -15 \leq \delta < 15 \right \} \\ 
\nonumber P_2 &= \left \{ \partial_x(x_0, y_0 + \delta, z_0) ~|  -15 \leq \delta < 15 \right \} \\
\nonumber P_3 &= \left \{ \partial_x(x_0, y_0, z_0 + \delta) ~|  -50 \leq \delta < 50 \right \} \\
\nonumber \vdots \\
\nonumber P_9 &= \left \{ \partial_z(x_0, y_0, z_0 + \delta) ~| -50 \leq \delta < 50 \right \} \\
\nonumber P_{\rm final} &= \left\{ P_1 \cup P_2 \cup \cdots \cup P_9 \right\} 
\end{align}
For convenience, we further down-sample $P_{\rm final}$ to 60 elements, which defines the feature vector. We determined the degree of downsampling that was appropriate by examining the widths of typical features in the derivative profiles by eye. Nevertheless, this smoothing may lead to worse performance.

\textbf{PCA}. We approximate the $30\times30\times100$ sub-cube around each pixel $p_i$ as a linear combination of 15 representative ``basis cubes''. We derive the basis cubes using principal component analysis (PCA, \citealt{Francis99}), and these basis cubes capture $\sim 95\%$ of the variance in the data. The 15 weights in the linear combination define our final feature vector for $p_i$. This is essentially a (lossy) compression of the data and, unlike the first two methods, does not explicitly encode any intuitive, identifying characteristics. Nevertheless, this expression of the data has proved useful in other classification tasks (e.g., asteroid taxonomy \citep{Tholen84}, stellar spectral types \citep{Singh98}, star/galaxy separation \citep{Cabanac02}). PCA has also been used to decompose and analyze molecular cloud structure \citep{Heyer97, Brunt09}.

\subsection{Training and Optimization}
\label{sec:optimize}
As discussed above, two free parameters influence the training process: $\gamma$ and $C$. We use cross-validation to choose optimal values for these parameters. We first partition our classification examples into two independent sets. The first (the training set) is used to train the classifier using a given value for ($C$, $\gamma$). We then apply the classifier to the second (validation) data set, and measure the accuracy of the identification. We repeat this process for different values of ($C$, $\gamma$) to maximize the performance on the validation set. This approach provides some protection against over-fitting, since over-fits to the training data will poorly classify the validation set. To maximize the independence of the training and data set, the two samples were drawn from different regions of the cube.

\section{Results and Discussion}
\subsection{Classification Performance}
\label{sec:performance}

 \begin{figure}[h!]
 \includegraphics[width=3.4in]{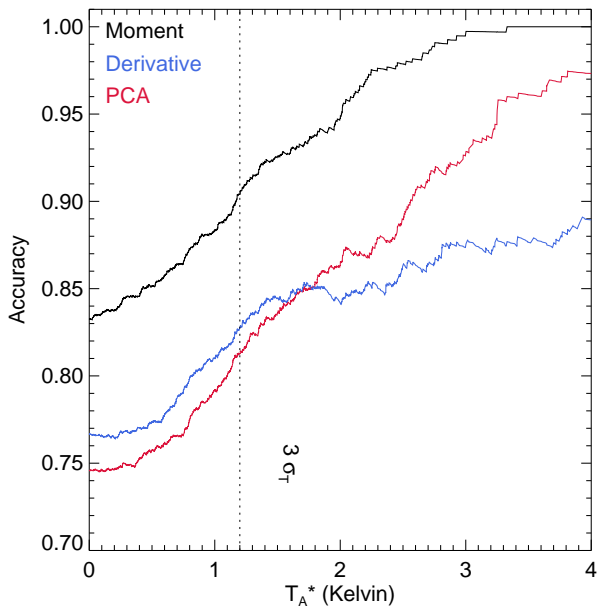}
 \caption{The performance of each SVM classifier as a function of intensity. We measure the accuracy of the classification by comparing to a manually-classified subset of the data (independent from the training data). The lines indicate the fraction of structures correctly classified, with brightness temperatures exceeding the value on the X axis. The three lines denote SVMs trained using different feature vectors, and the vertical line denotes the 3$\sigma$ noise level. All three classifiers are more accurate when classifying brighter structures.}
 \label{fig:optimize}
\end{figure}

\begin{figure}[h]
\includegraphics[width=3.4in]{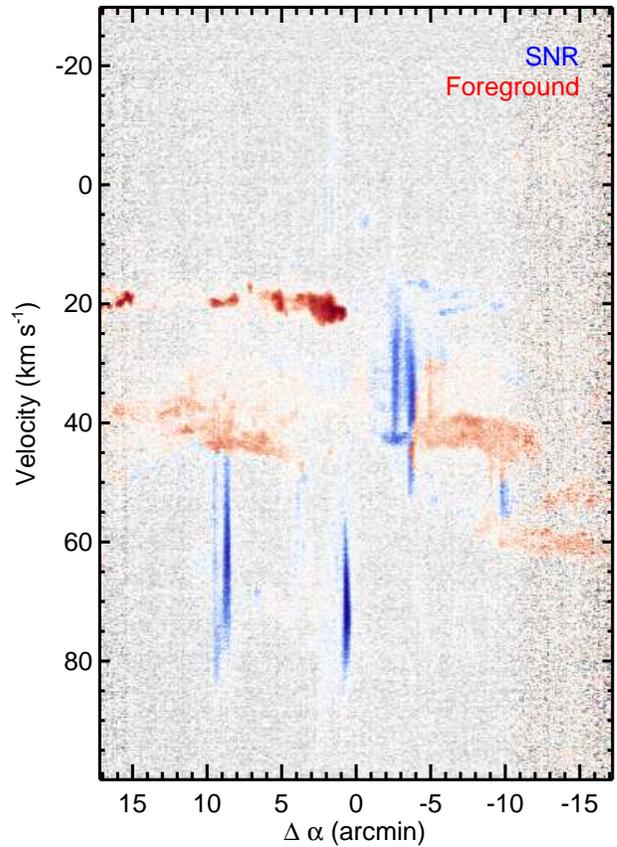}
\caption{The SVM classification of the same data shown in Figure \ref{fig:slice}. Blue pixels are classified as belonging to the supernova, while red pixels denote foreground cloud emission.}
\label{fig:slice_class}
\end{figure}

\begin{figure*}
\includegraphics[width=3in]{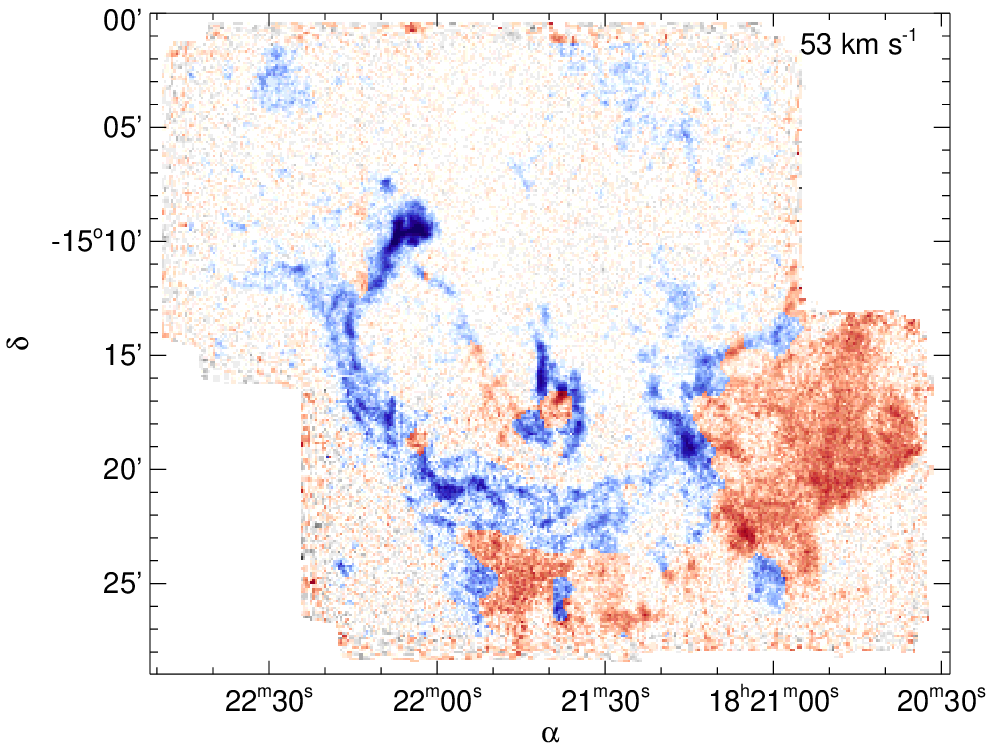}
\includegraphics[width=3in]{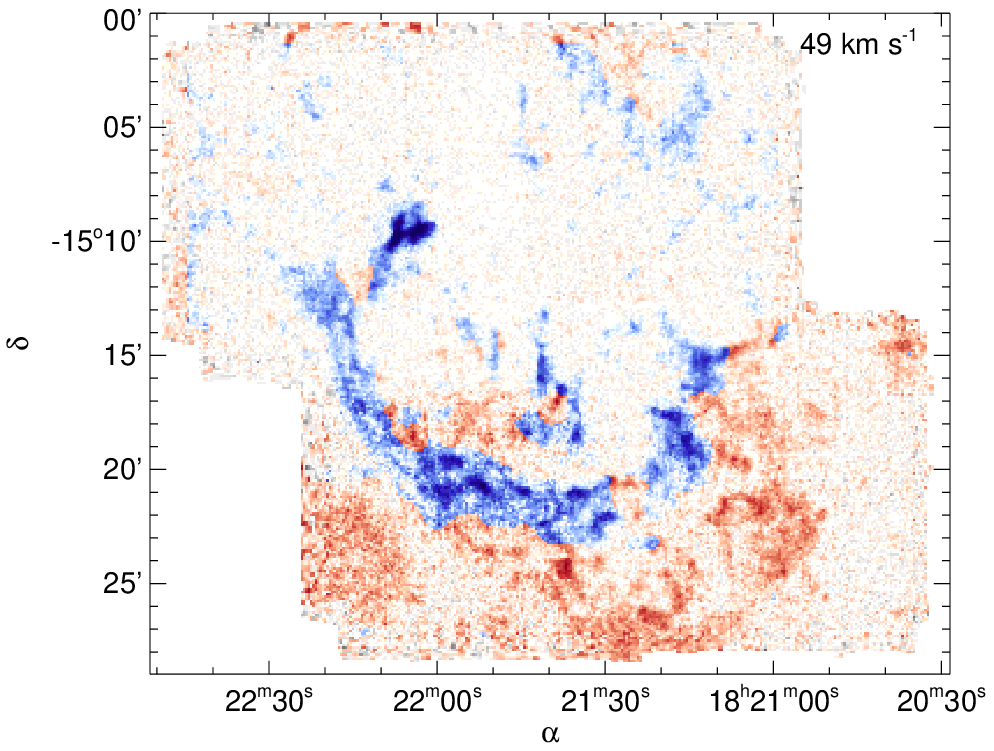}\newline
\includegraphics[width=3in]{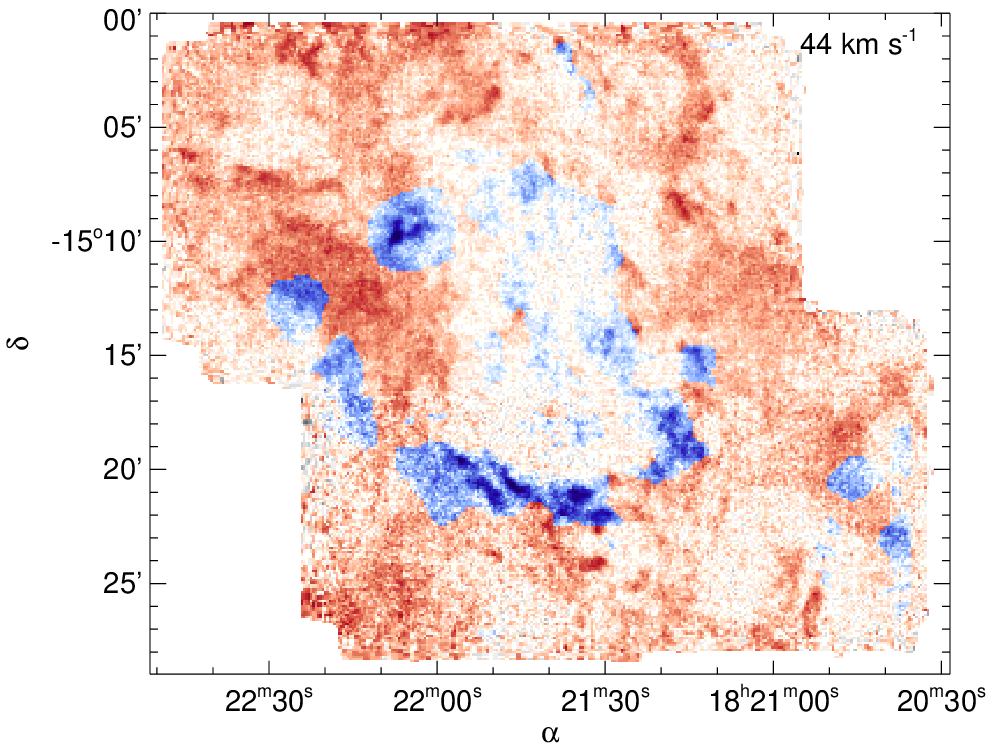}
\includegraphics[width=3in]{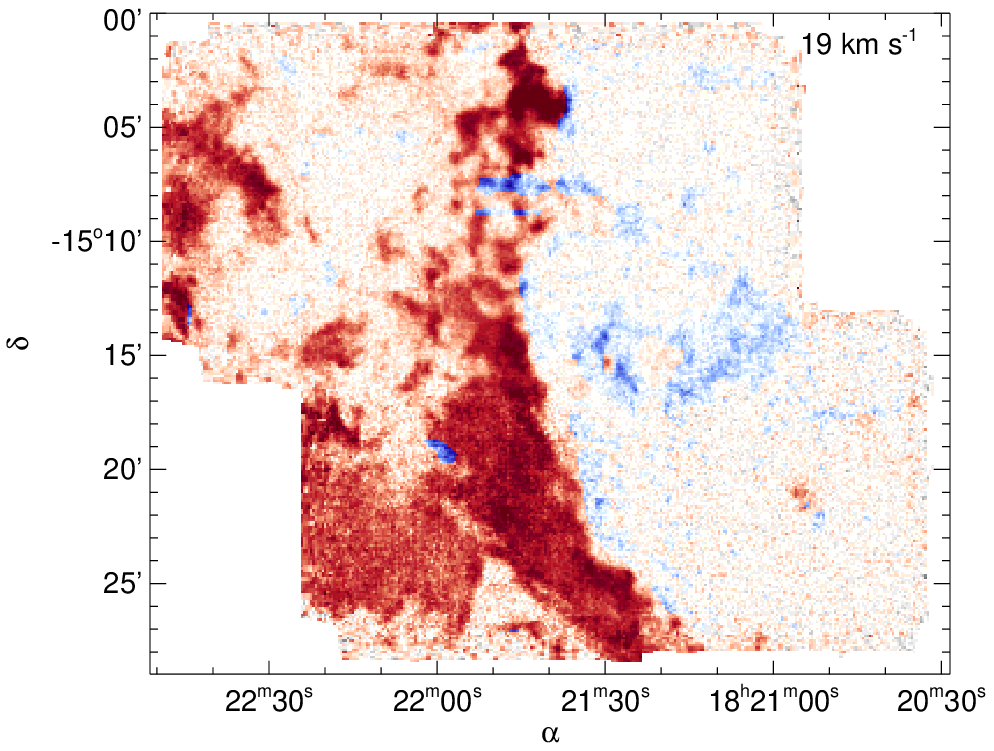}
\caption{The same as Figure \ref{fig:slice_class}, but for the position-position slices through the cube also shown in Figure \ref{fig:brogan}. An animation showing the full classification can be found in the electronic version of this article.}
\label{fig:pp_slice}
\end{figure*}

We evaluate the performance of each classifier by comparing the accuracy with which it classifies the validation data described in \S \ref{sec:optimize}. The accuracy is simply the fraction of correctly-labeled pixels. We find that the feature vector which encodes the moments of the intensity achieves the highest performance. 

The maximum accuracies achieved using the Moment, Derivative, and PCA feature vectors were 83\%, 77\% and 75\%, respectively (the y intercept of Figure \ref{fig:optimize}). However, misclassifications are biased towards low-intensity pixels. This is to be expected since, as Figure \ref{fig:slice} shows, pixels corresponding to blank sky have been included in both classes in the classification set. Hence, the proper classification of these faint pixels is not well defined. Misclassifying noise is not problematic, however, since simple thresholding later in the analysis can separate signal from background. For most data analysis purposes, it is more important that emission features be correctly classified. 

Figure \ref{fig:optimize} shows the accuracy at which each classifier identifies pixels above a given intensity threshold. Here, the moment-based classifier has an accuracy of 90\% for emission detected at 3$\sigma$, and exceeds 95\% accuracy for the brightest pixels in the data. Figure \ref{fig:slice_class} shows the classification of the data in Figure \ref{fig:slice}, using the Moment feature vectors. Figure \ref{fig:pp_slice} shows the classification of several position-position channels. Many of the misclassified bright pixels lie near the boundary between supernova and cloud emission. This is perhaps expected, since the moments of the intensity are measured (and implicitly smoothed) over a $30\times30\times100$ sampling window. Nevertheless, several supernova lines intersect and, presumably, extend behind cloud material. In most cases, the SVM classifier at least partially follows these transitions from supernova to cloud.

\begin{figure}
\includegraphics[width=3.2in]{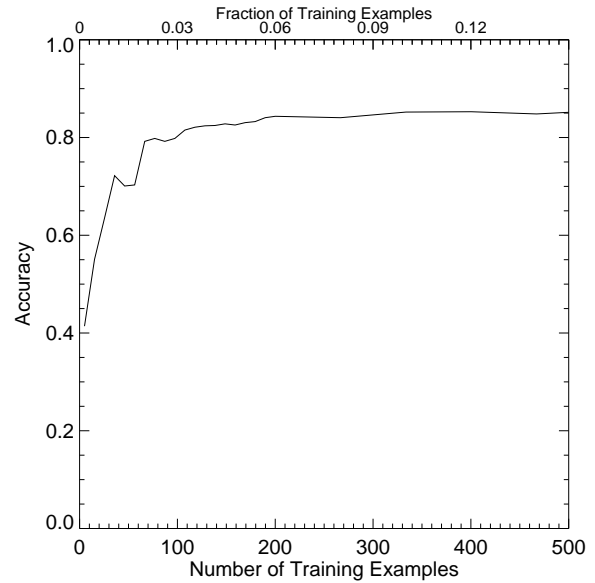}
\caption{The accuracy of the classifier using the Moment feature vector,  as a function of the training set size. The accuracy is measured using the entire data set (i.e., no filtering based on intensity). }
\label{fig:learning_rate}
\end{figure}

\begin{figure}
\includegraphics[width=3.2in]{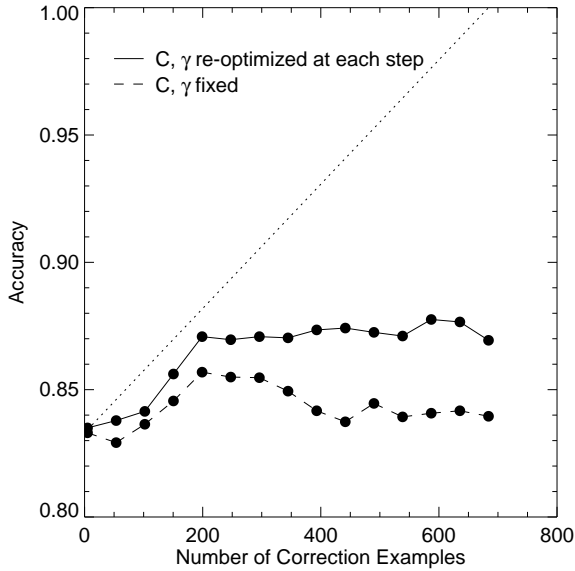}
\caption{The accuracy of the classifier using the Moment feature vector, as mistakes from the test data set are added to the training set. The solid line re-optimizes $C$ and $\gamma$ at each step. The dashed line fixes these parameters at their original optimal values. The dotted line is the trivial case of over-fitting, when each correction example causes the algorithm to correctly identify one additional example in the test set.}
\label{fig:correction_rate}
\end{figure}

There are three reasons why an SVM classifier mis-classifies data. First, data from two classes may not segregate perfectly in feature space (due either to noise in the data, errors in the training set, or a poorly-designed feature vector). Second,  the topology of the decision boundary (determined by the SVM kernel function) may not be able to conform to the distribution of training data in feature space. Finally, the training examples may insufficiently sample how data are distributed in feature space. Each of these possibilities has implications for how to design and improve classification pipelines. 

Figures \ref{fig:learning_rate} and \ref{fig:correction_rate} provide some insight into what limits the performance of our classification using the Moment feature vectors. Using the optimal values of $C$ and $\gamma$ found above, we measured the learning rate -- the accuracy of the classification as a function of training set size. Figure \ref{fig:learning_rate} shows this function, and suggests that most of the meaningful information is contained within the first few hundred examples. Remaining examples contain redundant information, and confer little performance gain.

Misclassifications in the validation set may occur because the training set isn't representative enough, and the validation set samples a systematically different region of feature space. To test this possibility, Figure \ref{fig:correction_rate} shows the classification accuracy when we re-train using the original training set, augmented with a subset of the validation data that were originally mis-classified. The solid line re-optimizes $C$ and $\gamma$ at each step. The dashed line shows the result when we fix these parameters to the values used above.

Note that in this experiment, part of the validation data is now explicitly included in the training set. This increases the risk of over-fitting the training data (i.e. devising arbitrary rules that fit the training data, but which do not generalize well to new data). As a trivial case of over-fitting, when a classifier is re-trained with a correction example in the validation set, it corrects the misclassification for that example only. The dotted line in Figure \ref{fig:correction_rate} depicts this scenario. Any meaningful performance gain should fall above this line, since the classifier should ideally use the information in each correction example to correct many additional mistakes.

The figure shows that, even when presented with additional correction examples, the algorithm shows essentially no performance gain. This figure rules out the possibility that misclassifications in the validation set are due to the training and validation sets sampling different regions of feature space. Instead, this strongly suggests that the classification task is limited by the partial overlap of the two classes in feature space. Additional training and correction examples are of little help in this situation, and a better feature vector is needed for further performance gain.

We do not claim that these classifiers will necessarily generalize to other data sets or classification tasks. Other applications likely require re-training the SVM algorithm using the data at hand, or testing new feature vectors. However, Figure \ref{fig:optimize} suggests that the SVM algorithm is capable of identifying morphological differences in the ISM, and Figure \ref{fig:learning_rate} implies that this task can be taught efficiently, with little manual classification. For example, we have started to investigate whether wind-blown bubbles can be identified using \textit{Spitzer} colors and edge information as a feature vector (Beaumont et al. in prep.)

\subsection{Mass and momentum of SNR G1605-0.57}
As mentioned above, the CO emission from \snr{} is likely due to the remnant's collision with a molecular cloud, presumably in the Scutum galactic arm at 3.4 kpc. Our pixel-level classification of the data allows us to analyze the properties of this emission in isolation from foreground material. Here we derive an estimate of the mass and momentum associated with the cloud/remnant collision. 

In the limit that all material along the line of sight can be described by a single excitation temperature $T_{ex}$, the equation for the observed radiation temperature $T_R$ is \citep{Ginsburg11}
\begin{equation}
T_R = T_0\left(\frac1{e^{T_0/T_{ex}}-1} - \frac1{e^{T_0/T_{CMB}} - 1} \right) \left(1-e^{-\tau} \right) f
\label{eq:trad}
\end{equation}
\noindent where $T_0 = h \nu / k = 16.6K$ for CO 3--2, $\tau$ is the optical depth, and $f$ is the beam filling factor (which we take to be 1). Furthermore, 
the column density of the J=3 state is given by
\begin{equation}
N_{J=3} = \frac{8 \pi \nu^3}{c^3 A_{32}} \frac1{e^{T_0/T_{ex}} - 1} \int \tau_v dv
\label{eq:ncol}
\end{equation}
\noindent where $A_{32}$ is the Einstein A coefficient. Equation \ref{eq:trad} can be solved for $\tau$, such that Equation \ref{eq:ncol} explicitly depends only on $T_{ex}$ and $T_R(v)$ (see, e.g., Equation A8 of Ginsburg et al. \citeyear{Ginsburg11}). 

\begin{figure}
\includegraphics[width=3.4in]{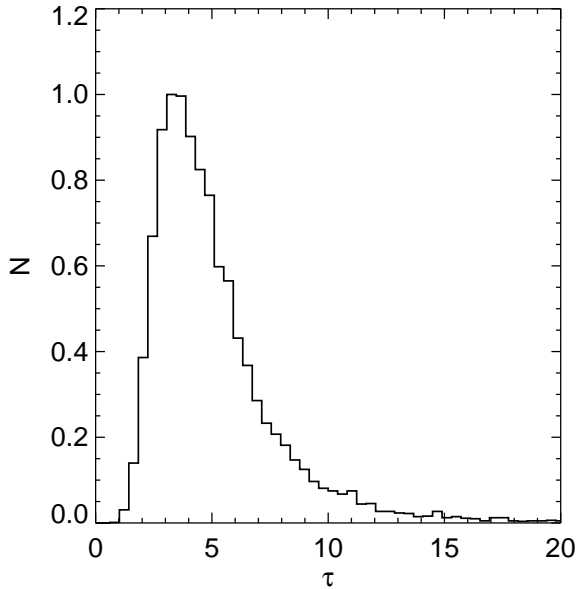}
\caption{Normalized histogram of the $^{12}$CO opacity inferred from Equation \ref{eq:tau}, in regions where both $^{13}$CO and $^{12}$CO are detected.}
\label{fig:tau}
\end{figure}

To constrain the excitation temperature and opacity in the line centers, we obtained supplementary $^{13}$CO J=3--2 observations towards three bright knots of emission. Assuming that the filling factor of the gas and excitation temperature of the two CO isotopologues are the same, their intensity ratio gives the gas opacity:
\begin{equation}
\frac{T_{12}}{T_{13}} = \frac{\nu_{12}}{\nu_{13}} \frac{ 1 - \text{exp}[-\tau_{12}]}{1 - \text{exp}[-\tau_{12} X_{13/12}]}
\label{eq:tau}
\end{equation}
\noindent where $X_{13/12}$ is the abundance ratio of $^{13}\text{CO}/ ^{12}\text{CO}$, which we take to be 70. Figure \ref{fig:tau} shows the inferred opacity for all pixels where we detect emission from both isotopes. The typical optical depth is 3--5. 

Plugging $\tau=4$ into Equation \ref{eq:trad} gives an estimate of the excitation temperature along each line of sight, which we find to be $T_{ex} = 15-30$K. This in turn allows us to evaluate Equation \ref{eq:ncol}. Finally, we convert from $N(J=3)$ to $N(CO)$ assuming the population levels are thermalized, and to $N(H_2)$ using an abundance ratio $X(CO) = N(CO)/N(H_2)= 10^{-4}$. The abundance of CO within shocks is uncertain. In their study of CO and H$_2$ vibrational lines in the C-type shocks of the Orion KL region, Watson et al. (\citeyear{Watson85}) measure an $X(CO)$ of $1.2 \times 10^{-4}$. On theoretical grounds, the value of $X(CO)$ for dissociative shocks may be enhanced by up to a factor of 100 if the re-formation of $H_2$ in post-shock gas is less efficient than CO \citep{vanDishoeck92}. Thus, the actual value of $X(CO)$ depends on both the type and strength of shocks in \snr{}, as well as the microphysics of grain catalysis in shocked gas.

\begin{figure}
\includegraphics[width=3.4in]{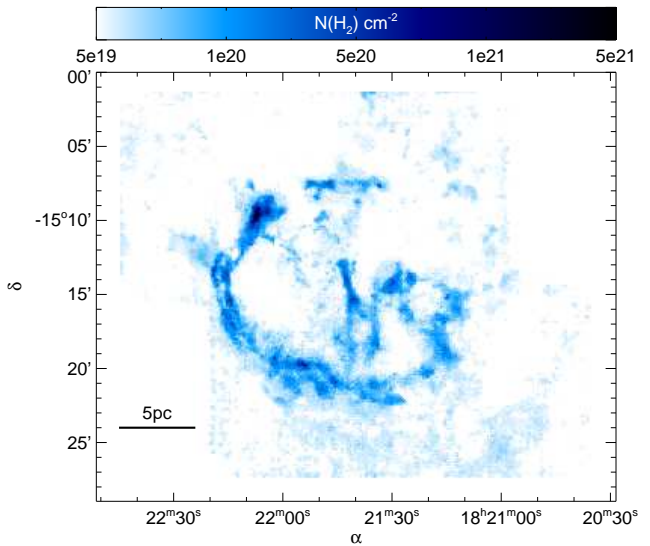}
\caption{The column density map of \snr{}. The scale bar assumes a distance of 3.4 kpc.}
\label{fig:ncol}
\end{figure}

Figure \ref{fig:ncol} presents the column density map of the supernova remnant obtained from this analysis. The angularly-integrated column density of the map is $\int\,N(H_2) d\Omega = 1.1\times10^{16} $\,cm$^{-2}$\,ster. If we further assume that the remnant is located within the Scutum arm at 3.4 kpc, this implies a total mass of $M = 2300$ M$_\odot$. We approximate the velocity of the gas at each point by the observed velocity dispersion -- this is a lower limit, since it only accounts for motion along our line of sight. Nevertheless, this implies a total momentum of $2.2 \times 10^4$ M$_\odot$ km s$^{-1}$. For comparison, the typical momentum of a supernova explosion is $\sim 1-10 M_\odot \times 10^4 \text{km s}^{-1} \sim 10^{4-5} M_\odot \text{km s}^{-1}$.

The characteristic width of filamentary features in the remnant is 30''--60'' = 0.5--1 pc. Taking this to be the line-of-sight depth of supernova emission implies a volume density of $n = N / l \sim 500-1000$ cm$^{-3}$. This is roughly 1--2 orders of magnitude lower than the densities van Dishoeck et al. (\citeyear{vanDishoeck93}) measured towards the supernova remnant / molecular cloud collision IC 443. The most likely explanation of this discrepancy is that the filamentary features in our data consist of unresolved filamentary substructure. If this is the case, then the characteristic depth $l$ would be smaller, and the corresponding volume density higher. However, this would not affect our mass and momentum measurements, since the increase in volume density is cancelled out by the decrease in solid angle.

\section{Conclusion}

We have presented a case study of a supervised learning task applied to classifying structures in the ISM. The M17 molecular cloud overlaps shocked CO emission from \snr{} but, because of the supernova's distinct morphology in position-position-velocity space, the two objects are readily distinguishable by eye. The SVM classification algorithm is able to learn these morphological differences using a representative sample of manually-classified pixels. We emphasize several important characteristics of this approach: 

\begin{enumerate}
\item Machine-based classification of ISM structures permits a pixel-level classification of datasets. This level of refinement is often prohibitively cumbersome via manual identification.
\item By using an independent set of manually-classified validation data, we can characterize the quality of this classification. We can further use this information to refine and improve the algorithm's performance.
\item Extracting information about the moments of the intensity distribution in our data produced the most effective classification. Other information (the weights in a principal component analysis, spatial derivatives) was less successful. 
\item Only a very small fraction of the data ($\sim0.1\%$) needs to be categorized to train the algorithm. An efficient approach for future work may be to evaluate the classifier's performance as the training set is assembled, to better understand when the training set is large (and representative) enough. 
 \end{enumerate}
 
This case study suggests that automated algorithms are capable of identifying complex structures seen in the ISM. Such an approach may be useful in analyses of current and future surveys of the Milky Way's ISM, particularly when identifying morphologically distinct structures like bubbles, pillars, and filaments.

We are grateful to C. Brogan for sharing her 20\,cm data of \snr{}, and to K. Binsted for discussions. We also thank the anonymous referee, whose careful reading and comments improved the clarity of this paper.  This material is based upon work supported by the National Science Foundation under Grant No. AST-0908159.
 
 \pagebreak

\bibliographystyle{apj}

%
%
\end{document}